\documentclass[acmart,acmsmall,authorversion]{acmart}
\AtBeginDocument{%
  \providecommand\BibTeX{{%
    \normalfont B\kern-0.5em{\scshape i\kern-0.25em b}\kern-0.8em\TeX}}}

\acmDOI{XXXXXXX.XXXXXXX}
\acmJournal{PACMHCI}
\acmYear{2025} \acmNumber{CSCW Preprint} \acmMonth{10} 

\usepackage{graphicx}
\usepackage{multirow}
\usepackage{subcaption}
\usepackage{xcolor}
\usepackage{titlesec}
\usepackage{tcolorbox}
\usepackage[normalem]{ulem}
\usepackage{enumitem}
\usepackage{tabularray}
\usepackage{caption}

\usepackage{tablefootnote}
\usepackage{booktabs}
\usepackage{vcell}
\usepackage{wrapfig}

\titleformat*{\subparagraph}{\bfseries}

\begin{document}

\title[]{An Experimental Study Of Netflix Use and the Effects of Autoplay on Watching Behaviors}

\author{Brennan Schaffner}
\affiliation{%
 \institution{University of Chicago}
\country{USA}
}
\author{Yaretzi Ulloa}
\affiliation{%
 \institution{University of Chicago}
\country{USA}
}
\author{Riya Sahni}
\affiliation{%
 \institution{University of Chicago}
\country{USA}
}
\author{Jiatong Li}
\affiliation{%
 \institution{University of Chicago}
\country{USA}
}
\author{Ava Kim Cohen}
\affiliation{%
 \institution{University of Chicago}
\country{USA}
}
\author{Natasha Messier}
\affiliation{%
 \institution{University of Chicago}
\country{USA}
}
\author{Lan Gao}
\affiliation{%
 \institution{University of Chicago}
\country{USA}
}
\author{Marshini Chetty}
\affiliation{%
 \institution{University of Chicago}
\country{USA}
}

\renewcommand{\shortauthors}{Schaffner et al.}

\newcommand{\mc}[1]{\textcolor{magenta}{MC: #1}}
\newcommand{\brennan}[1]{\textcolor{cyan}{B: #1}}
\newcommand{\lan}[1]{\textcolor{purple}{L: #1}}
\newcommand{\yaretzi}[1]{\textcolor{purple}{Y: #1}}
\newcommand{\riya}[1]{\textcolor{purple}{R: #1}}
\newcommand{\nat}[1]{\textcolor{purple}{N: #1}}
\newcommand{\ava}[1]{\textcolor{orange}{A: #1}}

\begin{abstract}
Prior work on dark patterns, or manipulative online interfaces, suggests they have potentially detrimental effects on user autonomy. Dark pattern features, like those designed for attention capture, can potentially extend platform sessions beyond what users would have otherwise intended. Existing research, however, has not formally measured the quantitative effects of these features on user engagement in subscription video-on-demand platforms (SVODs). 
In this work, we conducted an experimental study with 76 Netflix users in the US to analyze the impact of a specific attention capture feature, autoplay, on key viewing metrics. 
We found that disabling autoplay on Netflix significantly reduced key content consumption aggregates, including average daily watching and average session length, partly filling the evidentiary gap regarding the empirical effects of dark pattern interfaces. 
We paired the experimental analysis with users' perceptions of autoplay and their viewing behaviors, finding that participants were split on whether the effects of autoplay outweigh its benefits, albeit without knowledge of the study findings.
Our findings strengthen the broader argument that manipulative interface designs \textit{can and do} affect users in potentially damaging ways, highlighting the continued need for considering user well-being and varied preferences in interface design. 
\end{abstract}

\ccsdesc[500]{Human-centered computing~Empirical studies in HCI}

\keywords{attention capture, dark patterns, deceptive design, user agency, streaming media, autoplay, time management, right of access data requests}

\maketitle
\section{Introduction}
\label{sec:introduction}
The Human-Computer Interaction (HCI) community has seen growing interest in how online platforms use manipulative interface designs, often termed dark patterns or deceptive designs~\cite{brignull-2023}, to steer users toward decisions they would not otherwise make.
These patterns exploit cognitive biases and behavioral tendencies, often to the detriment of users. 
As platform interactions continue to play a central role in our daily lives, understanding the implications of dark patterns becomes crucial for ensuring ethical and user-centric design practices.

In particular, ``attention capture damaging patterns'' (ACDPs)~\cite{monge2023defining}, a subset of dark patterns designed to capture and retain user attention, have been the focus of several recent CHI studies~\cite{lukoff2018makes,lukoff2021design,bedjaoui2018user,schaffner2023don,monge2023defining,chaudhary2022you}. 
Employing such design elements to arrest user attention can contribute to stress, reduced productivity, and addiction-like behaviors, affecting users' mental well-being and overall health~\cite{chaudhary2022you, monge2023defining, schaffner2023don, cho2021reflect, lukoff2021design} by leading users to engage with content they might not have chosen willingly. Some researchers have specifically called out \textit{autoplay}---a platform design feature in which multimedia content, such as videos, is played one after another without explicit user consent---as a ubiquitous and especially harmful ACDP~\cite{lukoff2021design,monge2023defining}. The captivating nature of autoplay can undermine the agency of users' experience and morph their consumption patterns in ways they may regret~\cite{schaffner2023don}.

The backlash against autoplaying content extends beyond the academic realm. Mainstream journalism in the United States (US) has targeted the complications of autoplay, especially with respect to parenting children's device usage~\cite{nytimesautoplay, mashableautoplay, voxyoutuberabbithole}.
Recently, in the US, where large subscription video-on-demand (SVOD) platforms like Netflix are headquartered and widely used~\cite{statista_svod_countries}, bipartisan legislation has been drafted and introduced on multiple accounts which, if passed, would prohibit the use of autoplay (among other attention capture features) specifically for children (DETOUR Act~\cite{detouract}) or social media platforms (SMART Act~\cite{smartact}). To date, the European Union (EU)'s General Data Protection Regulation (GDPR) already prohibits subjecting users to automated decision-making without explicit consent~\cite{gdpr}.
Moreover, autoplay has also been labeled as an accessibility issue by independent accessibility organizations like the Bureau of Internet Accessibility due to the nature of media playing without consent being potentially distressing for neurodiverse users~\cite{boiaautoplayaccessibility, w3autoplayguidelines}. 

Yet, autoplay is extensively utilized by SVOD platforms (e.g., Netflix, Hulu, Disney+) who may argue that this feature is simply a user convenience~\cite{graypatterns}. Many argue to the contrary, that the platforms employ autoplay to benefit from the increased time that users spend on their platforms~\cite{monge2023defining,schaffner2023don,chaudhary2022you}, but these arguments are void of quantitative evidence.
That is, to the best of our knowledge, the impact of autoplay on SVOD users has not been quantified. Qualitative studies, such as diary studies~\cite{chaudhary2022you} and interview studies~\cite{schaffner2023don}, have shined a light on Netflix users' perceptions of autoplay but lack quantitative metrics.
We expand upon these works by asking: \begin{itemize}
    \item RQ1: To what extent (if any) does autoplay quantitatively affect users' video consumption? \item RQ2: What informs Netflix users' decisions to disable autoplay or keep it enabled?
    
\end{itemize}
    To this end, we employed a controlled experiment and systematic mixed-methods analysis, statistically analyzing quantitative effects of autoplay on users' watching behaviors on Netflix. We focused on Netflix, the original pioneer in the online TV streaming industry, because it remains the largest SVOD platform to date~\cite{netflix_vs_other_svods}. 
In September of 2022, over three-quarters of surveyed households in the US subscribed to Netflix~\cite{statista_netflix_households}.

Our controlled experiment included 76 moderate--heavy Netflix users in the US, split equally into treatment and control groups.
Participants in the treatment group turned off autoplay in their Netflix settings, while participants in the control group kept autoplay on. All participants installed a browser extension we developed to enable verification and monitoring of their study compliance by systematically sending participant information to our server back-end, including consistent logs of their autoplay settings. 
After a 10--17 day study period, we asked participants to exercise their rights to request their Netflix data---enabled by regulations~\cite{gdpr, ccpa}---to establish their extensive and accurate Netflix usage for both a baseline period leading up to the study and throughout the study itself. Participants submitted their anonymized data requests to the research team via our tailor-made web portal. 

We analyzed the participants' submitted data to look for differences in the participants' watching behaviors between the study's intervention period and the six-month duration leading up to the study, using the control group to account for confounding variables.
We found that disabling Netflix's autoplay resulted in, on average, 21 fewer minutes watched per day and 17-minute shorter sessions (consecutive viewings).
We also found that disabling autoplay increased the time between content viewings within a session by about 24 seconds on average but had no effect on the time between sessions.
Our findings suggest that by heightening the users' active role in playing Netflix content, more time is taken between episodes, users watch less overall, and sessions end sooner. 

Furthermore, we paired the experiment with pre- and post-study surveys, which we analyzed to further contextualize the experimental outcomes. 
We found that participants, prior to the study, generally felt in control of their time on Netflix and that the potential benefits realized from disabling autoplay may not outweigh its basic convenience for all participants. 

In summary, the primary contributions of this work are: 
\begin{itemize}    
    \item A first analysis of a dataset comprised of Netflix users' data requests, which provides both descriptive metrics of our participant population's general Netflix use and a baseline for statistical analysis. 
    \item A controlled experimental study measuring the quantitative effects of autoplay use on Netflix users' watching behaviors, supported by user perceptions from throughout the study. 
\end{itemize}

Our findings shed light on the substantial impact of autoplay on user engagement and the nuanced considerations that users may have regarding their experiences with autoplay.
In pursuit of tailored user-centric controls to cater to personalized content consumption needs while fostering a more empowering user experience, we recommend increased accessibility and control over autoplay features on SVOD platforms such as Netflix. 
 
\section{Background and Related Work}
\label{sec:related}
In this section, we provide an overview of relevant work in the fields of dark patterns, attention capture, and ``binge-watching''. We then provide a brief history of Netflix and a description of its current platform as the leading SVOD platform.

\begin{figure}
    \centering
    \includegraphics[width=\textwidth]{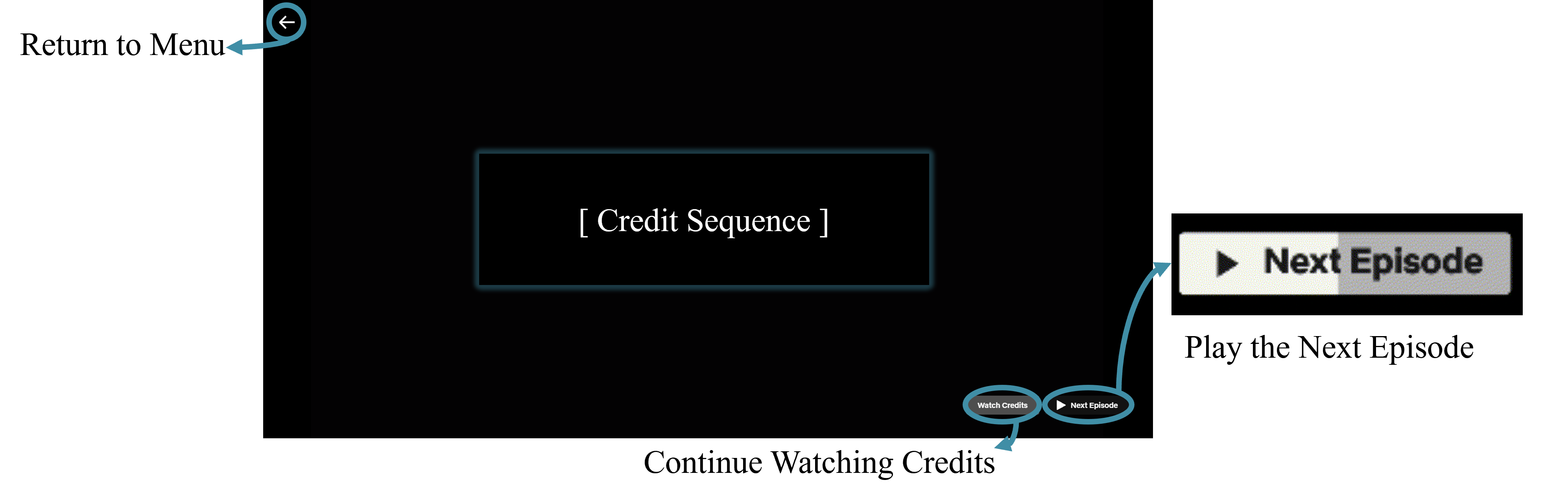}
    \caption{Illustration of how autoplay appears at the end of an episode on Netflix. Users may (1) return to the menu, (2) opt to continue watching the credit sequence, stopping the next episode from automatically playing, (3) opt to play the next episode right away, or (4) do nothing, for which the next episode will automatically play after about 5 seconds. A color wipe across the Next Episode button operates as a visual countdown timer for when the next episode will automatically begin.}
    \label{fig:autoplayexplain}
\end{figure}

\subsection{Dark Patterns in Streaming Media Platforms} 
Many online services design user interfaces (UI) to manipulate and exploit user behaviors and activities. These dark patterns in UIs have been widely recognized in the context of online purchasing, privacy-related decision-making, and navigating settings pages~\cite{gray2021dark,narayanan2020dark,mathur2019dark}. Previous works provide comprehensive taxonomies of dark patterns, revealing how prevalent dark patterns are and how they can compromise users’ interests from various perspectives~\cite{gray2018dark,di2020ui,mathur2021makes,mathur2019dark}. Bongard-Blanchy et al.\ also suggest that users can struggle to navigate against effects brought upon by dark patterns~\cite{bongard2021definitely}.

Among the many types of existing dark patterns, some designs termed “attention capture damaging patterns” (ACDPs)~\cite{lukoff2018makes,lukoff2021design,monge2023defining,bedjaoui2018user} prolong user engagement by distracting users, undermining their sense of agency, and inciting addictive behaviors. 
ACDPs can be detrimental to users’ physical and mental health~\cite{schaffner2023don,monge2023defining,chaudhary2022you}, increasing the chance of experiencing regret~\cite{cho2021reflect,chaudhary2022you} and undermining users’ trust in the platform~\cite{schaffner2023don}. 
Researchers have highlighted autoplay specifically as one of the most common and harmful ACDPs in streaming media platforms~\cite{lukoff2021design,monge2023defining}. Moreover, in one of the leading dark pattern taxonomies, Gray et al. place autoplay in a subcategory of “forced action” dark patterns, specifically those utilizing ``attention capture'' mechanisms~\cite{gray2024ontology}.

Moreover, autoplay is usually activated by default and can be hard to deactivate~\cite{mildner2021ethical}. Autoplay is believed to extend viewing sessions by seamlessly transitioning users from one piece of content to the next~\cite{cho2021reflect}, removing the need for any active decision-making by users~\cite{lukoff2021design,monge2022towards,burr2018analysis}, but without quantitative evidence to support this assertion. There is an effort in modern platform design to eliminate signals that suggest to users that it might be time to move onto a new activity (``stopping cues'')~\cite{alter2017irresistible}. The inertial nature of autoplay ices over stopping cues, resuming playback of new and engaging content quickly as the previous one ends.

Dark patterns research has increasingly paid attention to autoplay’s adverse effects.
For instance, Chaudhary et al. report that autoplay use increases mindless behavior, especially as fatigue sets in over prolonged sessions~\cite{chaudhary2022you}. Lukoff et al.\ also suggest that autoplay is the dark pattern most responsible for compromising users' sense of agency~\cite{lukoff2021design}. Worse, Hiniker et al. found that autoplay can also have increased detrimental effects on minors by reducing their autonomy and self-regulation, resulting in increased intervention from parents~\cite{hiniker2018coco}. Further, autoplay's prevalence accompanies---and, in part, enables---the rise in ``binge-watching'', a well-studied behavioral practice that carries its own range of adverse effects (e.g., addiction-like behaviors~\cite{riddle2018addictive, pierce2016just, devasagayam2014media}, detriments to physical health~\cite{exelmans2017binge, vizcaino2020tvs, spruance2017you}, and decreased mental well-being~\cite{de2016confessions,groshek2018netflix,ahmed2017new}). 
While there is academic consensus that autoplay encourages users to spend more time watching videos than they expected~\cite{lukoff2021design,schaffner2023don,monge2023defining,hanly2019switching,monge2022towards}, there is scarce evidence on how the use of autoplay quantitatively changes watching behaviors. 
Our work fills this gap with a statistical analysis of the changes in Netflix consumption when autoplay is disabled.

\subsection{From DVDs-by-Mail to Autoplaying Content} 
Founded in 1997, Netflix is now one of the most used SVOD platforms in the world~\cite{netflix_vs_other_svods}. Netflix was established in the US to provide DVD-selling and renting services ~\cite{reed2005}. They launched their service website in 1998~\cite{keating2012netflixed}, introduced their subscription model in 1999~\cite{Nast2002}, and delivered their billionth DVD by mail in 2007~\cite{billion}. By 2010, Netflix had evolved into the dominant SVOD service we recognize today, keeping an additional DVD subscription service on the side~\cite{pearson2011cult} and responding to the market's growing demand for predominately digital interactions. 

By 2016, Netflix's streaming service had expanded to almost every country worldwide. ~\cite{steel2016}. Today, Netflix has around 238 million paid subscribers worldwide~\cite{statista_world} and 75 million paid subscribers in the US and Canada~\cite{statista_US}. 
With the launch of Netflix Originals in 2013, the platform's growth was accompanied by a successful expansion into original content production, marking a shift from a content distributor to a producer of original content~\cite{jenner2016tviv, Carr2013}.
Matching societal trends of increased normalization and socialization of ``binge-watching'', Netflix's service has grown rapidly in scale and subscription.
This company and its service are now well-known for their largest market share of SVOD services worldwide~\cite{netflix_vs_other_svods}, attractive exclusive programming~\cite{statista_exclusive}, and refined content quality~\cite{statista_quality}. We focus our study on Netflix due to its prominence in the industry. 

A prominent feature of the contemporary Netflix experience is autoplay. For necessary study context, we present how Netflix's autoplay appears and functions in Figure~\ref{fig:autoplayexplain}.
When a piece of content playing on Netflix approaches its ending credits, a button appears in the bottom right, which users may press to begin playing the next piece of content (typically the next episode in a series). When autoplay is enabled---which is the default setting---the queued content will automatically begin to play after five seconds of inactivity by the user. 
When autoplay is disabled in one's Netflix account settings, the Next Episode button still appears, but the next episode is not played until the viewer chooses to do so. 
In 2020, Netflix added the ability for users to disable autoplay from their account settings found only on the browser version of the platform~\cite{jacobsnytautoplaycannowbeoff}, a feature which our study implementation takes advantage of, as detailed in the following section. 

Netflix's use of autoplay is similar to those employed by other SVODs (Disney+, Amazon Prime Video, \textit{etc.})~\cite{chaudhary2022you}. Like autoplay on Netflix, when one piece of content ends, a subsequent piece of content is automatically queued and starts playing if no user action is taken. 
Other short-form online video platforms like YouTube and TikTok also feature autoplaying content. Autoplay on YouTube operates much the same as on Netflix and other SVODs, and prior work found that autoplay on this platform limits users' sense of agency and causes them to feel less in control of their time spent on YouTube~\cite{lukoff2021design}. On the other hand, social media platforms that feature the automatic playing of video content, such as TikTok, employ a slightly different type of autoplay where user swipes could be 
considered a form of active consent. Still, work has experimentally determined that TikTok's platform design can reduce users’ ability to think analytically, increasing susceptibility to misinformation~\cite{jiang2024swiping}. The reduction in agency and analytical thinking resulting from platform designs that employ autoplay may lead to extended viewing sessions. In this paper, we test whether specifically autoplay on Netflix increases viewing time.

\begin{figure}
    \centering
    \includegraphics[width=0.9\textwidth]{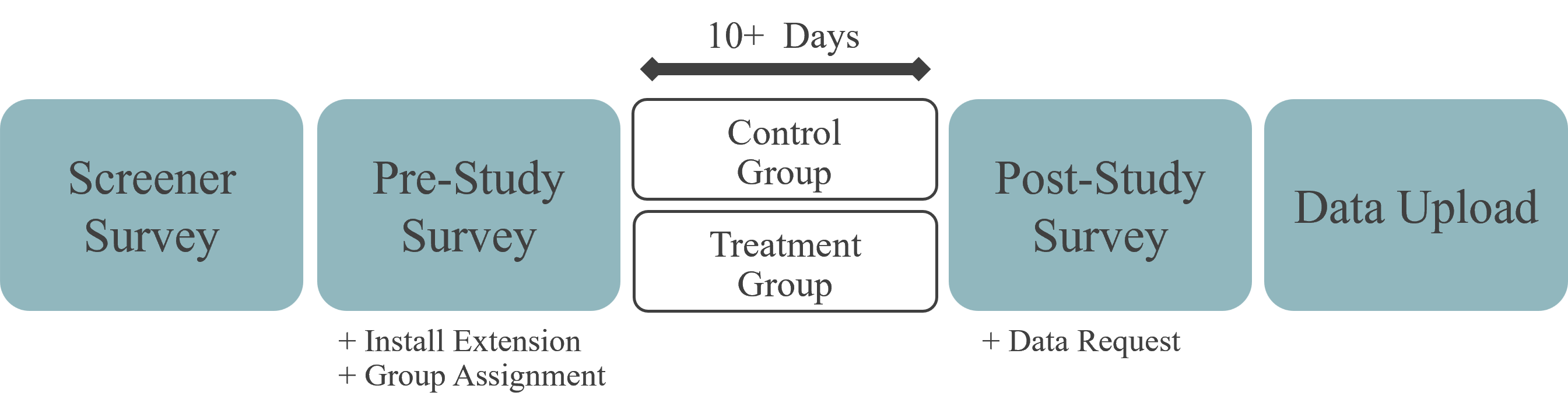}
    \caption{Structural components of the full study design, organized from left to right in the order that participants completed them.}
    \label{fig:studyparts}
\end{figure}

\section{Methodology}
\label{sec:methods}
To investigate the extent to which the autoplay feature on Netflix influences user watching behavior, we conducted a user study with the procedure illustrated in Figure~\ref{fig:studyparts} and detailed in this section. 
We developed all survey materials in Qualtrics~\cite{qualtrics} and deployed all study components using Prolific~\cite{noauthor_prolific_nodate}. Our Institutional Review Board (IRB) approved the study, and all materials are available in the Supplementary Materials. The study was conducted in ten separate, overlapping deployments from June to August 2023 to manage resource constraints.
\subsection{Participant Selection}
In this section, we describe the process for recruiting, screening, and enlisting participants, as well as the participant demographics.

\begin{table}[]
  \caption{Survey participant demographics: age, gender, highest level of education, estimated annual income, and timezone.} \label{tab:demographics}
  \resizebox{0.98\textwidth}{!}{%
  \begin{tabular}{lrr|lrr|lrr|lrr|lrr}
  \hline
  \textbf{Age} & \# & \% & \textbf{Gender} & \# & \% & \textbf{Education} & \# & \% & \textbf{Estimated Annual Income} & \# & \% & \textbf{Timezone} & \# & \% \\ \hline
  18-24 & 3 & 4 & Female Identifying & 37 & 49 & Less than High School & 1 & 1 & \$0-\$19,999 & 8 & 11 & EDT & 33 & 43 \\
  25-34 & 38 & 50 & Male Identifying & 39 & 51 & High School & 22 & 29 & \$20,000-\$49,999 & 18 & 24 & CDT & 29 & 38 \\
  35-44 & 20 & 26 & Non-binary & 0 & 0 & Associate's Degree & 6 & 8 & \$50,000-\$99,999 & 25 & 33 & MDT & 6 & 8 \\
  45-55 & 10 & 13 & Prefer not to answer & 0 & 0 & Bachelor's Degree & 32 & 42 & \$100,000+ & 12 & 16 & PDT & 8 & 11 \\
  55 years or older & 5 & 7 & Prefer to self-describe & 0 & 0 & Prof. or Advanced Degree & 15 & 20 & NA & 13 & 17 &  &  &  \\
   \hline
  \end{tabular}%
  }
  \end{table}

\begin{table}[]
  \centering
  \caption{Participants' reported their typical means of watching Netflix (participants could select multiple devices) and account ages, along with the number of other people using the same account and the percentage of time that participants watch by themselves (rather than with others).}
  \label{tab:account_info}
  \resizebox{0.9\linewidth}{!}{%
  \begin{tblr}{
    row{1} = {m},
    column{2} = {r},
    column{3} = {r},
    column{5} = {r},
    column{7} = {r},
    column{9} = {r},
    vline{4,6,8} = {-}{},
    hline{1-2,7} = {-}{},
  }
  \textbf{Device} & \textbf{\#} & \textbf{\%} & \textbf{Account Age} & \textbf{Years} & {\textbf{Number of Other People~}\\\textbf{Sharing the Same Account}} & \textbf{\#} & {\textbf{Portion that Participants}\\\textbf{Reported Watching Alone}} & \textbf{\%} \\
  Desktop/Laptop    & 52          & 68          & Median:              & 8.0            & Median:                                                               & 2.5         & Median:                                                                & 60.2        \\
  Mobile            & 50          & 66          & Average:             & 7.8            & Average:                                                              & 2.2         & Average:                                                               & 54.0        \\
  Smart TV          & 70          & 92          & Standard Dev.        & 4.3            & Standard Dev.                                                         & 1.9         & Standard Dev.                                                          & 32.5        \\
  Tablet            & 15          & 20          & Min, Max:            & (0.5, 18.0)    & Min, Max:                                                             & (0, 8)      & Min, Max:                                                              & (0, 100)    \\
  Gaming Console    & 15          & 20          &                      &                &                                                                       &             &                                                                        &             
  \end{tblr}
  }
  \end{table}
\subsubsection{Participant Recruitment}
We began by deploying a screening survey on Prolific to identify suitable candidates. We sought consistent, average-to-heavy Netflix users, which we considered to be individuals who reported watching Netflix at least three times a week, ranked it as their first or second SVOD platform, and did not frequently change their primary streaming service.
We also required that participants had used Netflix on their current account for at least six months to establish a sufficient baseline of watching behaviors.
Practical constraints required us to filter for US residents aged 18 or older who primarily used Google Chrome, which was necessary for the study's browser extension to monitor compliance. We also excluded individuals who had turned off autoplay in the past, previously requested their data from Netflix\footnote{Some platforms place restrictions on the amount and frequency for which its users can request personal data, and we did not want such limits to complicate the study.}, or did not have access to the email address associated with their Netflix account.

\subsubsection{Participant Screening}
The screening survey was completed by 1,203 individuals, 262 of whom met the criteria. These participants were compensated \$2 for completing the screener. We then invited 92 of these participants to the full study, with 78 completing all parts. After analyzing the data, we excluded 2 participants for misrepresenting their Netflix account activation period, leaving a final participant pool of 76. Each participant was compensated \$35 for completing all parts of the study, and those who dropped out were compensated for the components they completed, with their data excluded from analysis. 

\subsubsection{Participant Demographics}
Table~\ref{tab:demographics} shows the demographics of the 76 participants. 
While they spanned all age ranges, half were aged 25–34. The group included 39 males and 37 females, with 62\% holding at least a bachelor's degree. Most participants (81\%) lived in eastern or central US time zones. Additional characteristics, including Netflix usage patterns, are detailed in Table~\ref{tab:account_info}. Notably, 92\% of participants typically watched Netflix on a Smart TV, though they also used a variety of other devices. Participants had Netflix accounts ranging from 6 months to 18 years, with an average of about 8 years. They reported sharing their accounts with 0–8 other users, averaging about two users per account. Participants watched Netflix alone just over half the time on average, with considerable variability in individual preferences.

\subsubsection{Study Enrollment}
After screening participants and collecting their demographic information (see Figure \ref{fig:studyparts}), they completed a pre-study questionnaire through Prolific, were randomly assigned to either the \emph{control} or \emph{treatment} group, and received instructions for the study's technical setup. 
There were 38 participants in each group, with the \emph{control} group keeping autoplay on and the \emph{treatment} group having autoplay disabled.
The pre-study questionnaire provided qualitative context for their quantitative viewing data we discuss in the following sections.
For instance, the pre-study questionnaire asked participants to rate their agreement with statements regarding their sense of time management on Netflix and whether they liked to allow content on Netflix to play automatically.\footnote{To minimize participant priming, we resisted using the word ``autoplay'' in the pre-study questionnaire.}
Participants then installed a browser extension developed by the research team to track study compliance and followed instructions for configuring the extension and setting up their study conditions.

\subsection{Study Design}
This section describes the mechanics used for data collection. We outline how participant data was collected, anonymized, and submitted to the research team, as well as how accompanying qualitative data was collected with surveys.  

\subsubsection{Browser Extension}
We designed a Google Chrome extension to ensure participants followed the study’s technical requirements. The extension verified the autoplay setting—\textsl{off} for the treatment group and \textsl{on} for the control group—both at enrollment and throughout the study.
It sent logs tagged with each participant's unique ID to our \texttt{Google Cloud Storage Bucket}, with a \texttt{flask} backend running on \texttt{Google Cloud Run}~\cite{flask, googlecloud}.
The timestamped logs included any changes to a user's account autoplay settings and whether the extension was disabled prematurely. Since the Netflix account settings can only be accessed in the browser (and not in the app or on other devices like Smart TVs), the extension captures user attempts (or refusals) to change playback settings on the user's primary means of doing so---their computer browser. Also, since autoplay settings were configured in the users' Netflix account settings, the autoplay setting persists across all devices the users may watch on, not just the computer browser. For the sake of privacy, the extension was only active on necessary Netflix domains and did not track any other browsing activity.
Non-compliance resulted in data exclusion and termination from the study.

\subsubsection{Study Duration and Reflection}
Participants were instructed to watch Netflix as usual, with the \emph{control} group keeping autoplay on and the \emph{treatment} group turning it off. After 10 days, we sent a post-study survey through Prolific, including instructions for requesting account data from Netflix and reflection questions on their study experience.
For instance, we asked whether participants in the \emph{treatment} group, now having completed the study, planned to re-enable autoplay or keep it off and to provide any reasoning. 
Since participants took a variable amount of time to complete the post-study survey, the timeline that participants were enrolled in the study varied from 10--17 days. 

\subsubsection{Data Download} 
We leveraged participants’ Data Subject Access Requests (DSARs) to collect detailed Netflix usage data. This process is supported by privacy regulations like the EU’s General Data Protection Regulation (GDPR)\cite{gdpr} and California Consumer Privacy Act (CCPA)\cite{ccpa}.
Individuals can request access to their personal data by submitting a DSAR to the organizations (e.g., Netflix) that collect it. Most platforms offer DSARs in the form of downloadable files, which are sent to the individual via email within some set timeline of the request. 
DSARs from Netflix include detailed logs of all platform interactions, such as subscription history, device locations, and menu navigation. The most relevant data included for our study is each user's detailed watching history: titles of content watched, viewing times, and viewing durations. 

\subsubsection{Data Submission} 
For the final component of the study, we built a \texttt{Node.js}~\cite{nodejs} form hosted on \texttt{Railway}~\cite{railway} to collect participants' data requests.  
Since Netflix does not immediately process data requests, it was necessary to build this form separately from the post-study survey. From piloting, we found that two days was enough time for Netflix to process a data request, so we sent participants the upload form about 48 hours after they requested their data.

Prior to the study, the research team had analyzed their own Netflix data requests and were made aware of the potential privacy concerns that would come with sharing such data. 
For example, the data requests notably include the users' full names, email addresses, phone numbers, and locations. They also include expansive interactive histories like devices used, languages used, billing information, \textit{etc.} 
Critically, this information would also be included for those not enrolled in our study if they shared a Netflix account with someone who was enrolled; that is, data is included at the \emph{account} level rather than the \emph{profile} level. 

Consequently, to manage privacy concerns, our study’s data upload process is similar to those used in other user studies involving DSARs~\cite{wei2020twitter, borem2024data}. The web form walked participants through downloading the archive sent from Netflix and uploading it to our submission portal, which locally anonymized their data in-browser. Before sending the archive to the research team, the portal allowed participants to view the anonymized data but did not provide editing functionality.

We collected the following data from only the enrolled participant's profile: title of each piece of content viewed, watching duration, the date and time they watched it, and the country they were in when they watched it. We also recorded their Netflix subscription history to ensure baseline feasibility and participant honesty. 

\subsection{Data Analysis}
This section describes the methods for analyzing participants' Netflix data and survey responses.

\subsubsection{Data Preprocessing}
\label{sec:data_processing}
The participant DSARs include every video playback on the platform, regardless of type.
We categorized content from the Netflix data requests into \texttt{actual} (movies and TV series) and \texttt{promotional} (trailers, clips, automatically playing menu content, bonus interviews, \textit{etc}.) categories. 
We noted that this information is typically embedded into the title of the content in inconsistent forms (e.g., ``Season 1 Plot Clip: Rough Diamonds'' or ``Black Mirror: Season 6 (Episode Title Reveal)''). We used a combination of manual and automated methods to categorize titles as either \texttt{actual} or \texttt{promotional} content and further categorized \texttt{actual} content into either TV series or movies. 
We split the dataset into two parts based on the date of each participant's enrollment in the study (their baseline and study durations, respectively). Finally, we grouped participants' viewing histories into sessions, defining a \textit{session} as one or more consecutive viewings separated by less than 10 minutes. 
We then calculated time-based metrics and used statistical methods to compare the study’s results against baseline data.

\subsubsection{Baseline Analysis}
We used six months of pre-study viewing history as the baseline. For participants who had a break in Netflix usage, we extended the baseline period to ensure a full six months of data.

We analyzed the baseline dataset for descriptive metrics to characterize the participants' watching behaviors prior to the study. We calculate key time-based metrics such as \textsc{Average Session Length} to enable statistical analysis of the experimental effects as well as complementary metrics like the breakdown of time spent watching episodic TV content versus movies to provide context to the experimental outcomes.
To avoid skewing the data, we weighed each participant's data equally by first calculating the metrics for each participant's full baseline and then aggregating across participants. For numerical metrics (e.g., the number of unique movies watched), we report the median. For ranking metrics (e.g., most watched movies or most popular time of day to start a Netflix session), we report the result of a ranked vote (using the Borda count method~\cite{grandi2016borda}). 

\noindent\textbf{Pre-Study Survey Analysis}
Alongside the quantitative descriptions of the participants' watching behaviors, we also report the result of 5-point Likert-scale questions from the pre-study survey. The responses are grouped by theme and reported as the percentage of participants that selected each response (from Strongly and Somewhat Agree to Strongly and Somewhat Disagree).

\begin{table}
\centering
\caption{The four metrics we used to describe participant watching behaviors for the experimental analysis.}
\label{tab:metrics}
\resizebox{0.85\linewidth}{!}{%
\begin{tabular}{l|l} 
\hline
\textbf{Watching Metric}                                                & \textbf{Description}                                                                                                                                                                                                                                    \\ 
\hline
\textsc{Average Daily Watching}                                          & Average daily total watching time, in minutes.                                                                                                                                                                                                          \\ 
\hline
\textsc{Average Session Length}                                         & \begin{tabular}[c]{@{}l@{}}Average length of a Netflix viewing session, in minutes, where a session~\\includes subsequent viewings that begin within 10 minutes of the\\previous content's end. Time in between viewings is not included.\end{tabular}  \\

\hline
\textsc{Average Interim Time }                                          & The average time between viewings within the same session, in minutes.                                                                                                                                                                                  \\ 
\hline
\begin{tabular}[c]{@{}l@{}}\textsc{Average Time Between} \\\textsc{~Sessions}\end{tabular} & The average time between sessions, in minutes.                                                                                                                                                                                                          \\
\hline
\end{tabular}
}
\end{table}

\subsubsection{Experimental Analysis}
For measuring the effects of disabling autoplay, we focused on four key aggregates of participant watching behaviors, which are presented in Table~\ref{tab:metrics}. These four metrics together describe an encapsulating mosaic of the time participants spend watching Netflix. We compared these metrics for each group population between the study duration and baseline with the following method.

To make a fair comparison between the long six-month baseline and the participants' shorter participation in the study, we first calculated the duration for which each participant was enrolled in the study (10--17 days).\footnote{Some participants were provided study components, like the post-study survey, but did not complete them for a few days. If participants completed stages promptly, the shortest enrollment possible was 10 days, which was the case for most participants. Over 90\% of participants took just 10-12 days.} Then, we split up each participant's baseline into all possible contiguous periods of the same length. 
For example, if a participant was in the study for 11 days, we split up their 180-day baseline (six months) into 170 contiguous 11-day periods. 
We then constructed a series of ``baseline slices'' by taking one period from each participant, starting with the periods immediately prior to study enrollment and working backward until the baseline periods were depleted, keeping the control and treatment groups separate. Since the participant with the longest study timeline (17 days) resulted in the fewest number of baseline periods (164), we ended up with 164 feasible slices to use as the statistical baseline for the study. 
Constructing the ``baseline slices'' in this way ensures that each slice has the same composition as the study duration and the data used is as close in time to the study duration as possible. 

For each baseline slice, we calculated the four metrics for both the control and treatment groups by first calculating the metric for each participant and then taking the mean across participants within each group. 
We then represent the aggregated metrics from each baseline slice as a normal distribution so that one can locate where the respective metric from the study duration falls on the baseline distributions for both the control and treatment groups. Each study metric can be thought of as a Z-score~\cite{andrade2021z}, where the mean and standard deviation are calculated from the baseline distributions.
One would expect participation in the study to affect participant behavior (the Hawthorne effect~\cite{obrenovic2014hawthorne}). Thus, we must calculate the difference between the treatment and control groups' Z-scores to determine the effect of disabling autoplay while accounting for potential confounding factors such as the participant knowingly participating in a study about Netflix watching behavior. 

Finally, we posture the difference in Z-scores on a distribution of differences between two randomly sampled Z-scores ($\mu=0, \sigma=\sqrt{2}$), resulting in itself, another final Z-score that represents the probability that the difference between the two original Z-scores (from the treatment and control groups) is due to chance (i.e., a p-value). Moreover, since each Z-score, prior to taking the difference between the treatment and control groups, corresponds to a distance from the baseline mean in the same units as the metrics (minutes), we calculate the \textit{effect sizes} of disabling autoplay by combining the difference between the control group and its baseline with the difference between the treatment group and its baseline.

\noindent\textbf{Post-Study Survey Analysis}
Participants reflected on their study participation with both multiple-choice and open-ended responses in the post-study survey. We grouped participants by their multiple-choice responses and then analyzed the open-ended responses for common themes participants used to justify their multiple-choice responses. For example, we asked treatment group participants, having completed the study, whether they planned to re-enable autoplay, keep it disabled, or remain undecided. Then, we report percentage breakdowns for each choice, followed by typifying anecdotes enumerating the various reasons given.

\section{Findings} 
In this section, we first present the analysis of the participants' baseline Netflix consumption habits~(\S\ref{sec:baseline_findings}), which includes the most recent six months of participant watching data prior to the study along with pre-study survey responses. We then present the results of our experimental intervention study (\S\ref{sec:experiment_findings}), uncovering the quantitative effects of disabling autoplay on participant watching behavior. 

\subsection{Descriptions of Participants' Baseline Netflix Use}
\label{sec:baseline_findings}
We provide aggregate metrics and survey responses describing the participants' Netflix watching behaviors with respect to their baseline. That is, the following subsections summarize the participants watching behaviors and perceptions before participating in the experimental study.

\subsubsection{How much were participants watching?}
Here we describe what our dataset of the participants' baseline data indicates about \textbf{how much} the participants were watching in the time leading up to the study, including the aggregate metrics serving as a statistical baseline for the experiment and additional time-based descriptive analysis.

\begin{figure}
    \centering
        \captionof{table}{Baseline watching metrics aggregated across all participants. All four metrics are reported in minutes. }
    \label{tab:baseline_metrics}
    \resizebox{0.5\linewidth}{!}{%
    
      \begin{tabular}{l|c}
        \hline
        \textbf{Watching Metric} & \textbf{Baseline Aggregate} \\ 
        \hline
        Average Daily Watching  & \begin{tabular}[c]{@{}c@{}}$mean=51.3 $\\$SD=5.87$\end{tabular} \\ 
        \hline
        Average Session Length & \begin{tabular}[c]{@{}c@{}}$mean=55.9$\\$SD=4.33$\end{tabular} \\ 
        \hline
        Average Interim Time  & \begin{tabular}[c]{@{}c@{}}$mean=0.959 $\\$SD=0.076$\end{tabular} \\ 
        \hline
        Average Time Between Sessions & \begin{tabular}[c]{@{}c@{}}$mean=1100 $\\$SD=116$\end{tabular} \\
        \hline
      \end{tabular}

    }
\end{figure}

\textbf{Participants' Baseline Watching Behavior Leading up to the Study: }  
Table~\ref{tab:baseline_metrics} contains descriptive metrics for the participants' watching behaviors during the baseline period. On any given day, participants watched an average of about 50 minutes. Continuous viewing sessions lasted just under an hour on average (55.9 minutes). 
The time between consecutive viewings within a session was about 1 minute. (If the ``interim'' time was 10 minutes or longer, we considered this the start of a new session.) This is longer than the interim time provided by autoplay (about 5 seconds) because interim time can include time spent maneuvering the platform menus or taking short breaks. Finally, the average time between consecutive sessions was about 1100 minutes (18.3 hours). These metrics---the ones in Table~\ref{tab:baseline_metrics}---serve as reference points for the participant population prior to the quantitative differences uncovered from statistical analysis resulting from the experimental intervention (turning off autoplay). For additional context, the median participants typically watched about 7,440 minutes ($\approx$24 hours) of content across about 138 separate sessions spanning their six-month baselines.

\subsubsection{What kinds of content were participants watching?} 
We describe what our dataset of the participants' baseline data indicates about \textbf{what} the participants were watching in the time leading up to the study, including breakdowns of the types of content (TV series vs. movies), the relative popularity of specific titles, and their relationships to \texttt{promotional} content. In addition to exploring the types of Netflix content viewed in the participants' baseline, we also looked for any connections between autoplay and the content viewed by participants during the study. 

\textbf{Mostly Episodic Content: }
Participants watched TV series far more than movies. About 86\% of time spent viewing content on Netflix was for TV series. Moreover, TV series make up nearly 93\% of total viewing occurrences (rather than time spent), which is consistent with the fact that movie viewings, while fewer in number, last longer on average than a single TV viewing.
The dominance of TV series provides an interesting baseline for the experimental intervention since autoplay is especially applicable to the notion of consecutive episode viewings. 

\begin{table}
\centering
\caption{The top titles in our participants' baseline viewing, resulting from ranking titles by total viewing time and aggregating with Borda scoring. Since the top titles are all TV series, we also report the top movies separately.}
\label{tab:top10}
\resizebox{0.8\linewidth}{!}{%
\begin{tblr}{
  row{1} = {c},
  cell{1}{1} = {c=2}{},
  cell{1}{3} = {c=2}{},
  vline{2} = {1}{},
  vline{3} = {2-6}{},
  hline{1,7} = {-}{0.08em},
  hline{2} = {-}{},
}
\textbf{Top 10 Titles} &                     & \textbf{Top 10 Movies}     &                                   \\
1. Love is Blind       & 6. Shadow and Bone  & 1. You People              & 6. All Quiet on the Western Front \\
2. You                 & 7. The Night Agent  & 2. Anna Nicole Smith:[...] & 7. Luther: The Fallen Sun         \\
3. Perfect Match       & 8. Ginny \& Georgia & 3. A Man Called Otto       & 8. The Hunger Games: [...]        \\
4. Manifest            & 9. Black Mirror     & 4. Glass Onion:[...]       & 9. Missing                        \\
5. Wednesday           & 10. BEEF            & 5. Your Place or Mine      & 10. I See you                     
\end{tblr}
}
\end{table}

\textbf{Convergence of Top Titles: }
While participants' baseline together included a breadth of viewing activity (about 26,000 viewing instances totaling about 682,000 viewing minutes), certain titles clearly emerged as most popular. By median, each participant viewed about 23 unique TV series\footnote{We counted multiple seasons of the same series as viewings of the same unique TV series.} and 12 unique movies. We ranked each title by total time spent viewing and aggregated across participants using Borda scoring. Table~\ref{tab:top10} shows our participants' most watched titles. 
Notably, all of the listed titles reached Netflix's own reported Top 10 in the US at some point during the study baseline, indicating a level of agreement between our dataset and the broader US Netflix population~\cite{transparentdatareportnetflixcnn, netflixtop10records}. 

Even though Netflix's Top 10 titles emerged as the most popular across the six-month baseline, the Top 10 titles represented quite a small portion of the baseline viewing.\footnote{The average Jaccard similarity coefficient between a participant's baseline viewing and Netflix's Top 10 titles throughout the respective baseline period was about $0.017$.} Consequently, we could not conduct sufficiently powerful statistical tests to analyze any relationships between whether a participant had autoplay on or off and their consumption of Netflix's Top 10 content.

\textbf{High Exposure to Promotional Content: }
In addition to the viewing of TV series and movies discussed above, participants also viewed a high amount of \texttt{promotional} content. Recall from \S\ref{sec:data_processing} that we distinguish \texttt{promotional} content from \texttt{actual} TV series or movies to encompass all other forms of viewing (trailers, autoplaying banners on the landing page, teasers played from hovering over a selection, \textit{etc}.). The relationships between \texttt{actual} and \texttt{promotional} content viewed are meaningful. \texttt{Promotional} content made up less than 2\% of a participant's time spent on the platform but almost half (46\%) of all viewing instances.
This means that platform users encounter a high number of short-lasting \texttt{promotional} content throughout a Netflix session. Specifically, our participants encountered about 2--3 instances of \texttt{promotional} content per session, each lasting about 13 seconds on average. 

Though making up just a small portion of overall time on the platform, the high amounts of \texttt{promotional} content still serve as a major gateway to longer lasting \texttt{actual} content. Whenever a participant started viewing a new unique title (i.e., started an entirely new show or watched a different movie), in over half of such cases (60\%), the participant had previously viewed \texttt{promotional} content for that title. Autoplay on Netflix will frequently automatically play \texttt{promotional} content, especially upon completing a movie or the last episode available for a TV series. The high exposure to autoplayed \texttt{promotional} content paired with the high conversion rate of \texttt{promotional} content leading to \texttt{actual} content work in tandem to prolong viewing sessions and increase time spent on the platform.

\begin{figure}
    \centering
    \includegraphics[width=0.95\textwidth]{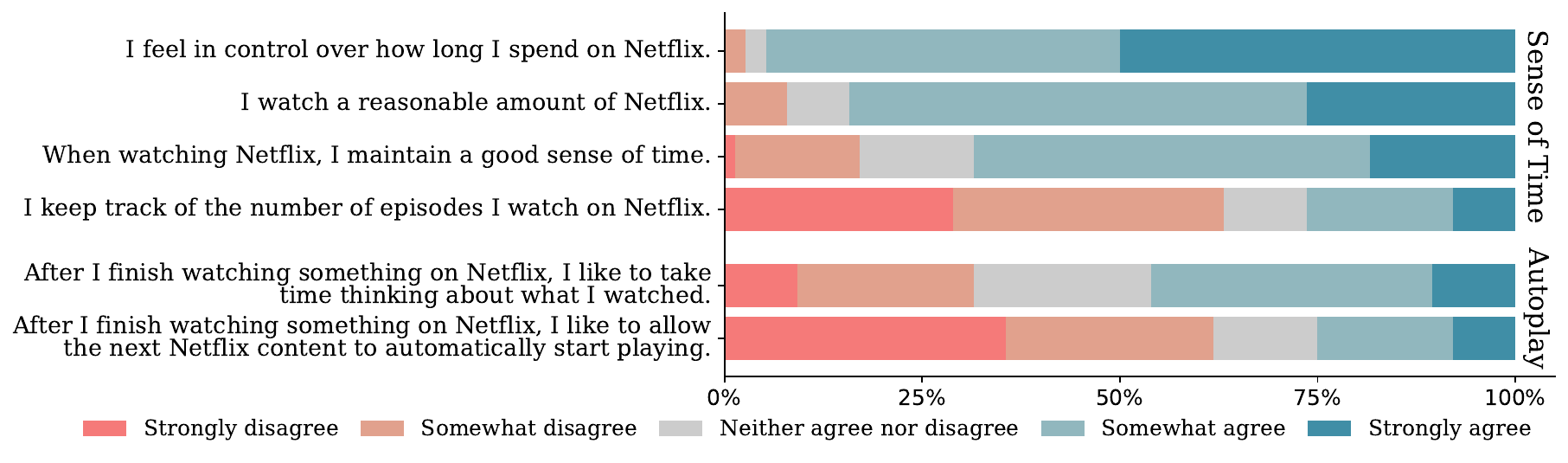}
    \caption{Likert responses regarding perceptions of watching time and autoplay.}
    \label{fig:timelikerts}
\end{figure}

\subsubsection{Prior to the study, how did participants regard autoplay and their time on Netflix?}
\label{sec:preregard}
We present the following user perceptions to provide further context to the experimental results and enrich the ensuing discussion. In particular, we present how participants responded to Likert-scale questions in a pre-study questionnaire designed to uncover participants' initial perceptions of their platform usage. 

\textbf{Participants Hold a Sense of Control: } 
Participants generally reported high rates of perceived agency and control over their Netflix watching behaviors, as shown in Figure \ref{fig:timelikerts}.
The overwhelming majority of participants felt in control of how long they spent on Netflix (95\% Somewhat or Strongly Agreed). Participants also felt that they watched a reasonable amount (84.2\% Somewhat or Strongly Agreed),  with moderately fewer participants reporting that they maintained a good sense of time when watching (68.4\% Somewhat or Strongly Agreed). 
While participants generally felt in control and content with their consumption amounts, only about 1 out of 6 participants kept track of the number of episodes they watched (18.5\% Somewhat or Strongly Agreed), potentially indicating that participants do not feel that episode tracking is important for managing their time and that they may employ other time-tracking strategies. 

\textbf{Autoplay May Counter Intentions: }
There is a relatively balanced split among participants about whether they have a propensity for reflective thinking after viewing content on Netflix: 46\% Somewhat or Strongly Agreed, while 32\% Somewhat or Strongly Disagreed. Participants showed more agreement when considering the value of content playing automatically.
A substantial proportion (62\% Strongly or Somewhat Disagreed) indicated a reluctance toward automatic content continuation, even though they had autoplay enabled. 
Autoplay is enabled by default, and previous work indicates that many Netflix users may not know that autoplay can be disabled at all~\cite{schaffner2023don}. 
Some participants' latent dissatisfaction with autoplay may be warranted, as the following section shows a reduction in overall consumption when autoplay is disabled. 

\textbf{Integration of Autoplay into Daily Lives: }
To garner additional insights about participants' baseline Netflix usage prior to the study, we asked if there was anything else participants would like us to know about their use of the platform in a free-response format. While most participants (57/76) had nothing else to add, the remaining quarter reported a variety of use cases. From playing Pup Academy for their dog (P39) and keeping a physical notebook to keep track of watched episodes (P2) to spending time with family (P9, P12, P55, P67), SVODs like Netflix have become a tool that accents the lives of users in a variety of ways. The respondents commonly reported using Netflix for background noise while conducting other activities like gaming (P15) or crocheting (P61) as well as relaxing and falling asleep (P3, P10, P14, P47, P52, P70). These observations highlight how Netflix usage is seamlessly woven into users' low-stakes daily routines, potentially amplifying the effect of platform features like autoplay and increasing overall time spent on the platform.

\begin{table}
\centering
\caption{Results from analyzing the participants viewing metrics from the experimental study.}
\label{zresults}
\resizebox{0.9\linewidth}{!}{%
\begin{tabular}{l|rl} 
\hline
\textbf{Metric}                        & \textbf{Result}                                                                                                     & \multicolumn{1}{c}{\textbf{Interpretation}}                                                                                                                                                     \\ 
\hline
\textsc{Average Daily Watching}        & \begin{tabular}[c]{@{}r@{}}\textit{p} = 0.003**\\ \textit{effect size} = (-)21.0 minute\end{tabular}                & \begin{tabular}[c]{@{}l@{}}Turning off autoplay resulted in about a 21 minute \\decrease in average watching time per day.\end{tabular} 
                        \\ 
\hline
\textsc{Average Session Length}        & \begin{tabular}[c]{@{}r@{}}\textit{p} = 0.013*\\ \textit{effect size} = (-)17.6 minutes\end{tabular}                & \begin{tabular}[c]{@{}l@{}}Turning off autoplay resulted in about an 18 minute\\ decrease in average session lengths.\end{tabular}  
\\ 
\hline
\textsc{Average Interim Time}          & \begin{tabular}[c]{@{}r@{}}\textit{p} = 0.0009** \\\textit{effect size} = (+)0.407 minutes\end{tabular}            & \begin{tabular}[c]{@{}l@{}}Turning off autoplay resulted in about a 24 second~\\increase in time between viewings within a session.\end{tabular}                              \\ 
\hline
\textsc{Average Time Between Sessions} & \begin{tabular}[c]{@{}r@{}}\textit{p} = 0.697 \\\textit{effect size} = N/A\end{tabular}                             & \begin{tabular}[c]{@{}l@{}}Turning off autoplay had no significant effect on \\the amount of time between consecutive sessions. \end{tabular}                                 \\
\hline
\end{tabular}
}
\end{table}

\begin{figure}[]
    \centering
        \begin{minipage}{0.48\textwidth}
        \centering
        \includegraphics[width=\linewidth]{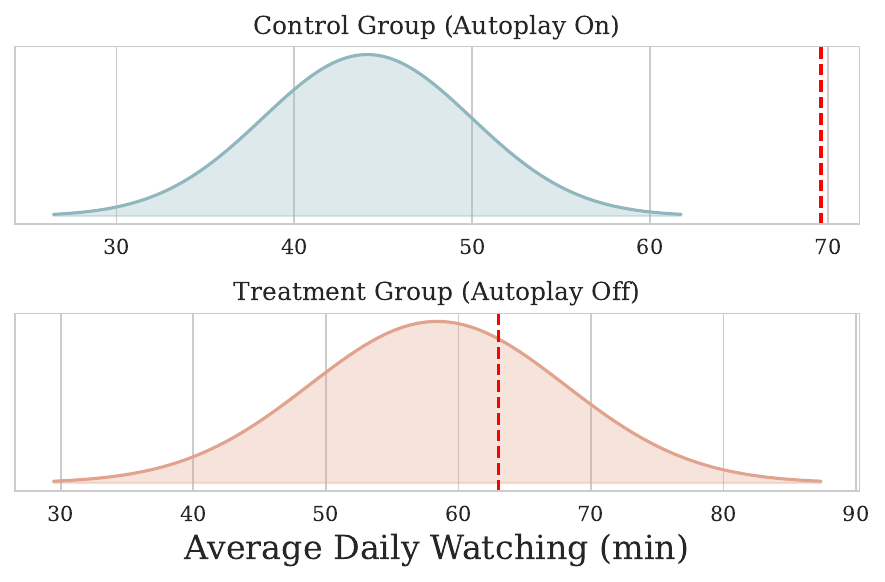}
        \caption{The curves represent the distribution of the Average Daily Watching metric across the six-month baseline split by control and treatment group. The vertical red line indicates the Average Daily Watching during the 10+-day study.}
        \label{fig:dist_dailywatching}
    \end{minipage}
    \hspace{0.01\textwidth}
    \begin{minipage}{0.48\linewidth}
        \centering
        \includegraphics[width=\linewidth]{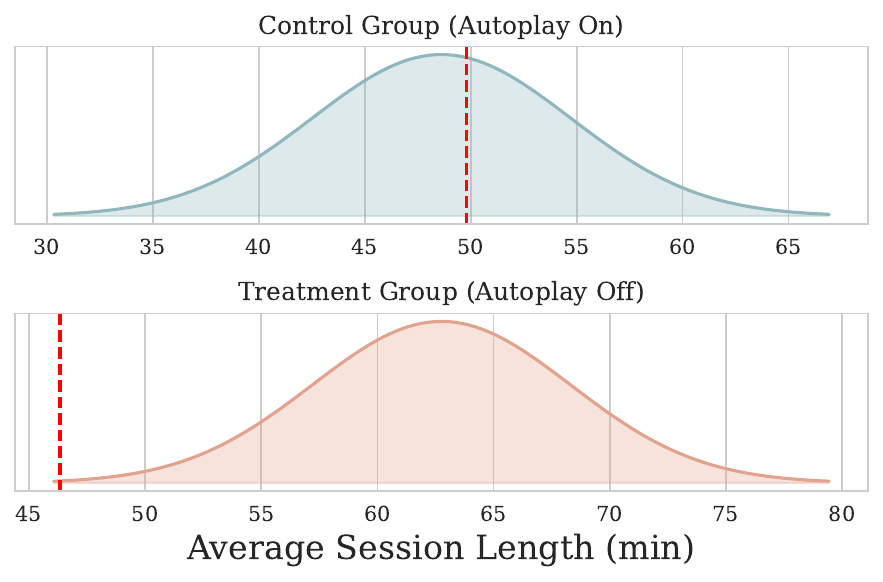}
        \caption{The curves represent the distribution of the Average Session Length metric across the six-month baseline split by control and treatment group. The vertical red line indicates the Average Session Length during the 10+-day study.}
        \label{fig:dist_sessionlength}
    \end{minipage}
\end{figure}

\subsection{Experimental Outcomes and Participant Reflections} \label{sec:experiment_findings}
Here we present the experimental outcomes. Notably, when participants in the treatment group (N=38/76) disabled autoplay, they exhibited substantial reductions in \textsc{Average Daily Watching} and \textsc{Average Session Length}, as well as substantial increases in \textsc{Average Interim Time}, compared to the control group (N=38/76). However, \textsc{Average Time Between Sessions} remained statistically inconclusive. We also present participants' reflections on having autoplay disabled for the study and how the effects of autoplay can be weighed against its convenience.

\subsubsection{Turning off autoplay reduced overall watching.}
Analysis from our experimental study indicates a notable reduction in total Netflix consumption from turning off autoplay when controlling for study effects. The quantitative experimental results are summarized in Table~\ref{zresults} and expanded upon below.

\begin{figure}[]
    \centering
        \begin{minipage}{0.48\linewidth}
        \centering
        \includegraphics[width=\linewidth]{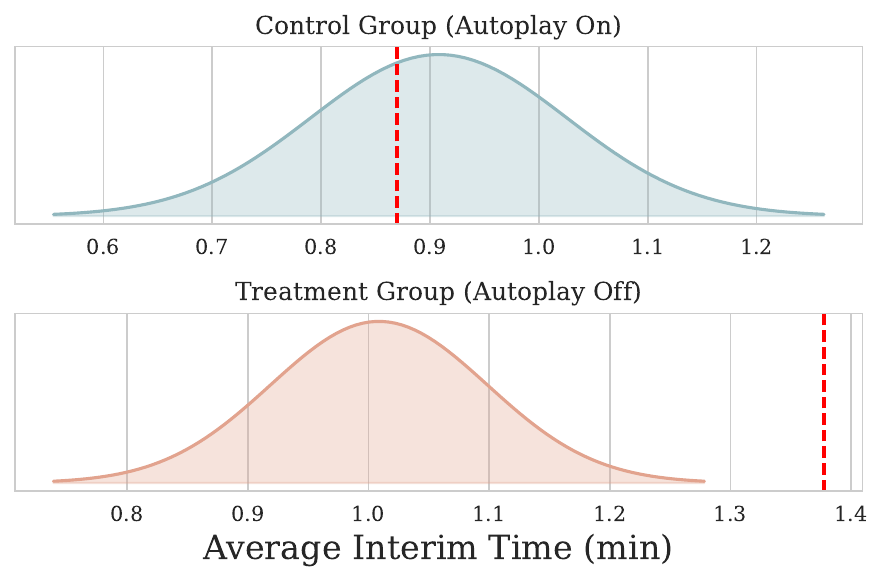}
        \caption{The curves represent the distribution of the Average Interim Time metric across the six-month baseline split by control and treatment group. The vertical red line indicates the Average Interim Time during the 10+-day study.}
        \label{fig:dist_interim}
    \end{minipage}
    \hspace{0.01\textwidth}
    \begin{minipage}{0.48\linewidth}
        \includegraphics[width=\linewidth]{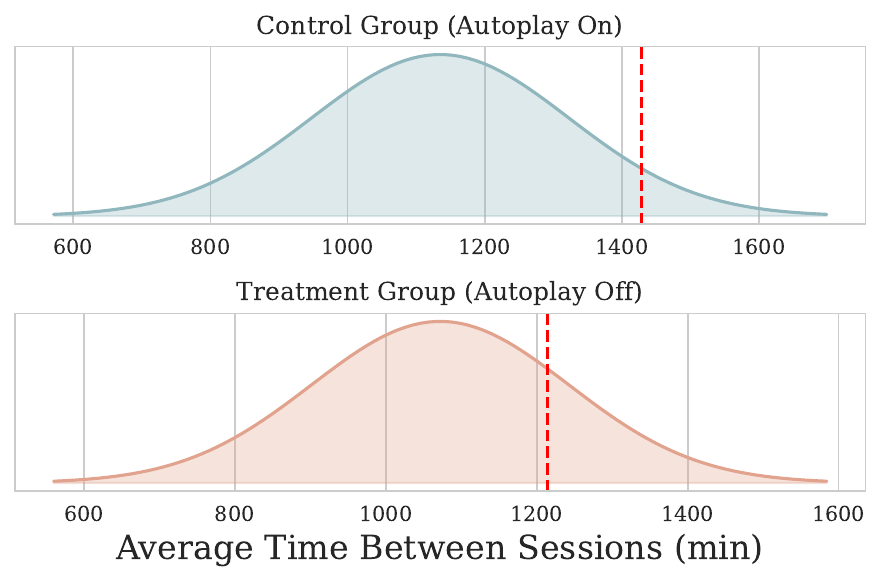}
        \caption{The curves represent the distribution of the Average Time Between Sessions metric across the six-month baseline split by control and treatment group. The vertical red line indicates the Average Time Between Sessions during the 10+-day study.}
        \label{fig:dist_betweensessions}
    \end{minipage}
\end{figure}

\paragraph{\bf 21.0 Minute Reduction in Average Daily Watching: } 
As shown in Figure~\ref{fig:dist_dailywatching}, the daily total viewing increased for both the treatment and control groups compared to their respective baselines. However, the treatment group's average daily watching time increased by significantly less than that of the control group. The fact that both the control and treatment group's metrics for average daily watching time increased can be explained by \textit{The Hawthorne Effect}, where participant behavior is modified under the awareness of being studied~\cite{obrenovic2014hawthorne}. Not only could participants infer that we were interested in their watching behavior, but they were also primed to think about Netflix more often, leading to increased visits to the platform.
This change would have been experienced similarly for both the treatment and control groups, but the disabling of autoplay appears to have, in part, counteracted the effect for the treatment group. 

\paragraph{\bf 17.6 Minute Reduction in Average Session Length: }
As shown in Figure~\ref{fig:dist_sessionlength}, the control group's average session length during the study was approximately the same as their average session length during the six-month baseline. However, the treatment group's average session length dropped significantly compared to their baseline, indicating that participants' viewing sessions were prolonged by autoplay. 
The p-value of 0.013 can be understood as the probability that one would see a difference of this size or greater between the control and treatment groups when randomly sampling one point from each of their respective baseline distributions.

\paragraph{\bf 24.4 Second Increase in Average Interim Time: }
Figure~\ref{fig:dist_interim} shows how the participants in the control group exhibited similar interim time as during their baseline. 
However, as one might expect, the participants in the treatment group---who disabled autoplay---exhibited longer downtimes between subsequent viewings within a session. 
Since autoplay did not automatically play the next piece of content, the extra time is likely due to the additional action required for participants to actively do so.

\paragraph{\bf No Significant Change in Time Between Sessions: }
We also analyzed the effects of autoplay on the average interim time \emph{between} sessions.
However, we found no evidence that time between sessions was affected. As shown in Figure~\ref{fig:dist_betweensessions}, the difference between how the control and treatment groups placed within their respective baselines was statistically insignificant. That is, if one were to randomly sample points from each group's baseline distribution, their respective Z-scores would exhibit a similar difference about 70\% of the time. 
As such, while disabling autoplay can cut sessions short and reduce overall watching, the effects may not be seen outside of sessions.

\begin{figure}
    \centering
    \includegraphics[width=0.9\linewidth]{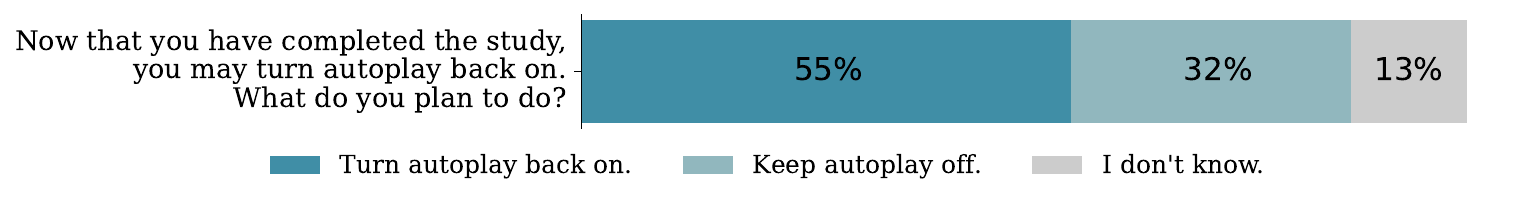}
    \caption{Participant responses from when we asked the treatment group (the 38/76 participants who disabled autoplay) whether they plan to turn autoplay back on having completed the study.}
    \label{fig:turnbackon}
\end{figure}

\subsubsection{When reflecting on study participation, perceptions of autoplay diverged within the treatment group.}
We asked the participants who disabled autoplay to reflect on their experience in the study and whether they would turn autoplay back on or keep it disabled. 

\textbf{Participants Consider Autoplay's Convenience: }
About half of the treatment group reported that they plan to turn autoplay back on upon having completed the study (Figure~\ref{fig:turnbackon}). These participants commonly referred to the feature's convenience, as highlighted by these anecdotes:  
\begin{quote}
    \textit{``I like it automatically going to the next episode \textbf{so I don't have to move out of my comfortable bed} and from what I was doing to find the remote'' (P49).\\
    ``I will end the episodes when I want to, even in the middle of one, so \textbf{autoplay being on just allows me to stay in a comfortable position} if I am watching on my computer or TV'' (P20).}
\end{quote}
Other reasons for turning autoplay back in included a preference for returning to normalcy. To this end, participants said they would re-enable autoplay because: 
\begin{quote}
    \textit{``I like putting things back \textbf{like they were}'' (P53).\\
    ``\textbf{It's what I've had for awhile} and I don't see any reason to keep it off'' (P50).}
\end{quote}
Netflix, by default, has autoplay enabled, and users are accustomed to its nature. If users do not feel that autoplay affects them negatively, it is reasonable for them to default to the platform's status quo. 

\textbf{Participants Considered Autoplay's Removal of Friction: }
About one-third of participants reported that they plan to keep autoplay disabled. 
Their responses commonly referenced the added ``friction'' required to continue watching Netflix once autoplay is disabled. P11 wrote to this end: 
\begin{quote}
\textit{``It was interesting, I didnt think about it at first, but then I realized \textbf{I had to grab my [remote] in order to move to the next episode}'' (P11).}
\end{quote}
Some participants further connected the increased friction to creating additional space for actively deciding whether to continue watching: 
\begin{quote}
\textit{``\textbf{It made me more conscientious of how many episodes I was watching} and if I was cleaning or up doing something, I would often not play the next episode'' (P47).}
\end{quote}
Further, P11 noted how the increased involvement between episodes served as mental waypoints when reflecting the length of the current viewing session: 
\begin{quote}
\textit{``It did make me realize how many episodes I was watching more so then as before. I didnt pay attention to it as much because it was automatic before. \textbf{Now I was like...`oh ok i had to do this 3 times so this is the 3rd episode}' '' (P11).}
\end{quote}
Overall, the added friction and increased opportunity for ending viewing sessions earlier likely explain the reductions in watching metrics. 

The remaining 13\% of the treatment group found themselves undecided on their autoplay preferences moving forward, grappling with the dilemma of balancing convenience and autonomy. A few participants encapsulated this struggle, remarking: 
\begin{quote}
    \textit{``On one hand, \textbf{I felt it was annoying} that it didn't move on to the next episode if I was doing something, \textbf{but it also helped me} to be in silence some and track my viewing better'' (P47). \\
     ``\textbf{I'm not sure yet.} I'll give it a few more days and see'' (P41).}
\end{quote}
This indecision mirrors the overarching quandary facing users, torn between the convenience offered by autoplay and the desire for agency over their viewing.

\subsection{Limitations}
There are several limitations to our study design. 
First, we confined our study to include only participants living in the US, restricting diversity in participants’ cultures and demographics. Future work could investigate whether these findings are consistent in other regions. Second, we use the participants' systemized watching logs to calculate their watching metrics. However, there is no way to distinguish whether users are actually watching the video content or if their attention is elsewhere (e.g., if they left the room or fell asleep). 
Still, our method captures how much video they \textit{played} on Netflix, which is a reasonable proxy for participant viewing, especially as it is implemented consistently across study duration and participation. 

Third, relying on self-reported data comes with the possibility of deviations from true behavior. We minimized this limitation by using the participants' real data logs collected by Netflix. However, participants were not monitored when downloading their Netflix data and uploading the data requests to our submission portal. While unmonitored, participants could have, in theory, taken extra steps to edit their data by unzipping the Netflix archive, editing the comma-separated values (CSV) file, and then re-zipping the altered files before uploading to our submission portal. 
We view this limitation as similar to issues in all studies that use self-reported data. For instance, users can lie in interviews and surveys which is also difficult to detect. 

Fourth, 10--17 days is relatively short in relation to the lifetime of a Netflix account. While we implemented a control group and statistical methods to account for sample variation, additional studies would be needed to confirm effects lasting longer than the study duration. 

Fifth, our method for classifying titles into TV series versus movies and \texttt{promotional} versus \texttt{actual} content was imperfect. We verified a random sample of 800 participant viewings and found 31 classification errors. Upon addressing the issues, we resampled another 800 viewings and found just 4 errors. %
In some cases, the title identifier provided by Netflix in users' data requests was entirely resistant to categorization (e.g., $6 Underground\_VAR2\-RECIPE\-3\_EDITOR2$ and $Placeholder\_SummerStreamingLM2023\ '23\ Trailer$).
While this was a negligible size of our dataset, it points to a need for platforms to continue to whole-heartedly abide by privacy laws by making data requests not only available but usable and clearly understandable. 

Finally, our browser extension monitored participants' autoplay status at the point where it is typically accessed: their browser. Recall that autoplay cannot be changed from the Netflix app (on Smart TVs, mobile devices, gaming consoles, \textit{etc}.). However, participants could have, in theory, disabled autoplay from their non-primary browser. We implemented several checks to minimize this risk. For instance, we checked that participants used the same Chrome browser---the one with the extension enabled---for all parts of the study even without us prompting them to do so. The browser extensions also sent logs whenever participants disabled it during the study, which we confirmed was never the case. We also ensured participants understood the study's compliance requirements as they both pledged in the pre-study (and confirmed in the post-study) that their Netflix account playback settings would be (and were) unchanged during the study. We monitored the extension logs closely for exceptions and had to discontinue some participants for not disabling autoplay in the first place when we asked them to configure their playback settings in the pre-study. 

\section{Discussion}
\label{sec:discussion}
In this section, we explore the multifaceted implications of turning off autoplay on the use of SVOD platforms, from fostering conscious decisions and potentially shorter sessions to revealing viewing habits for insightful decision support. 
We also explore the state of relevant regulations surrounding autoplay features. 

\subsection{Forcing Conscious Decisions}
Recently, dark patterns researchers~\cite{gray2024ontology, lukoff2021design, monge2023defining} have flagged autoplay as a potentially concerning feature due to its ability to undermine users’ autonomy. For example, prior work has equated autoplay to boarding a bus over which passengers have limited agency to determine where the stops are or when to get off~\cite{schaffner2023don}.
As users continue to rely on autoplay, they can lose their sense of time, inadvertently extending viewing sessions due to the loss of stopping cues and interaction friction. 
Our findings reveal that participants who watched content with autoplay enabled experienced increased overall watching time and moved between pieces of content more quickly. In contrast, the participants with autoplay disabled showed increased downtime between consecutive viewings,  allowing more time for making conscious decisions.
Disabling autoplay allowed participants to take advantage of their gained opportunities for reflection and make decisions that better aligned with their viewing goals, rather than being sped to more content or promotional material that pushes content.
These findings corroborate the core arguments made by many dark patterns researchers that ACDP interface designs capture and manipulate user attention.
Further research could build upon our work by continuing to pinpoint and quantify the influences of other ACDPs, such as infinite scroll.

\subsection{A Cost for Convenience}
Netflix advertises autoplay as a feature enabling convenience, which---while true---is not the whole story.
Our findings suggest that autoplay’s removal of friction can measurably contribute to the adverse effects brought on by over-consumption, such as disrupted sleep and other adverse health consequences~\cite{starosta2020understanding}.
SVODs need to more carefully weigh their responsibility in creating ethical platform experiences to better align with user desires and societal health.

We contribute to the ongoing discussion regarding the ethical considerations of autoplaying content. %
Beyond the dark patterns research that examines autoplay's role in undermining users' control over their time, researchers have targeted autoplay on other platforms---not SVODs---for its role in user radicalization~\cite{markmann2021youtube, giansiracusa2021autoplay} and forcing violent content on users without their consent~\cite{davisson2017autoplaying, jones2017murder}. While some participants appreciated autoplay's effects, the mentioned convenience may not be free.
Researchers have argued that autoplay increases platform marketability for advertisers, and thus, ad revenue is the true driver of platforms incorporating autoplay, not user convenience~\cite{davisson2017autoplaying}. 
Further, the innocuous integration of platforms like Netflix into users' lives and the casual standard at which users hold SVOD platforms may partially hide the effects of autoplay.  
By quantifying the effects of Netflix's autoplay in this work, we shine additional light on the hidden costs of convenience and underscore the need to critically reevaluate how platforms design for (un)ethical user engagement.
\subsection{Revisiting and Reimagining Autoplay Controls}
Netflix added autoplay to its platform in 2016 and then added the ability to turn it off in 2020~\cite{jacobsnytautoplaycannowbeoff}. Still, many users remain unaware of the ability to disable autoplay~\cite{schaffner2023don}. SVODs could commit to having autoplay disabled by default, requiring users to make an active choice when opting into the feature.
While most of the treatment group participants informed us that they planned to turn autoplay back on, about one-third opted to keep autoplay off after experiencing the benefits of disabling autoplay during the study. Users exhibit bespoke behaviors and preferences---an HCI truism.
To accommodate all users, we recommend that Netflix and similar SVOD platforms implement more accessible and heightened controls over autoplay settings. In doing so, platforms would allow users to select their own autoplay-specific preferences rather than asserting difficult-to-access default playback settings that become \textit{de facto} standards of the user experience. 

Custom alterations of autoplay across different users and between unique sessions for individual users are necessary to match the variable nature of user agency and flexibility.
For instance, users could specify the number of episodes autoplay can queue before being disabled for the remainder of that viewing session. 
The current implementation does not allow convenient customization outside of toggling the feature on or off. 
Previous research has criticized the brevity of the five-second autoplay countdown, arguing that it fails to provide adequate time for users to process content~\cite{chaudhary2022you}. 
To this end, the platform could allow users to configure the countdown duration. Creating longer countdown options would let users simultaneously experience the benefits of autoplay’s convenience while providing themselves adequate time to make conscious decisions. SVODs could even guide users toward revisiting their current settings after some time. Paired with heightened configurability, prompting users to reconsider autoplay settings could promote greater self-regulation and well-being amongst users by periodically adapting their preferences to their unique viewing habits.

\subsection{Revealing Viewing Habits: Insights and Decision Support}
Prior to study enrollment, there was a prevailing sentiment that participants were content with their viewing habits (\S~\ref{sec:preregard}), raising questions about the significance of advocating for the disabling of autoplay. However, as the study unfolded, several participants' reflective anecdotes highlighted newly realized areas for enhancing their relationships with the Netflix platform, suggesting that there might be untapped opportunities for platforms to better align with users' evolving preferences and needs. 
Moreover, participants may have been unaware of the influence of autoplay on viewing behaviors; the outcomes of this research could serve as a potential catalyst for broader public awareness. While respecting users' autonomy in their choices, these findings offer quantitative support that might inform and guide users' viewing decisions. Looking ahead, future work could build upon these results and delve deeper into discrepancies between users' perceived and actual viewing behaviors. 
\subsection{Autoplay Regulation}
As our study shows, the impact of autoplay is palpable for adults in the US, raising concerns about potential heightened susceptibility among children.
There is initial regulatory support targeting design features such as autoplay, with proposed laws in the US targeting autoplay features for kids~\cite{detouract} and on social media platforms~\cite{smartact}. 
This work prompts a crucial consideration for future research to examine whether ACDPs have exacerbated effects on children. 
The dynamics of children watching autoplaying content become more complex when parents face conflicting roles, balancing promoting autonomy with ensuring well-being.
The Federal Trade Commission (FTC)'s fair practice laws (Section 5 of the FTC Act) include prohibitions for unfair or deceptive practices affecting commerce including harm to consumers~\cite{ftcsection5}. The Commission evaluates conduct based on various criteria, including the existence and avoidability of harm while also considering public interest and awareness. Considering that platforms are designed for compulsive use, which is linked to addiction-like behavior~\cite{alter2017irresistible}, and given that children watching Netflix may struggle to avoid autoplay, it becomes imperative to explore ways to address this issue.
Furthermore, the European Union's General Data Protection Regulation (GDPR) already mandates that users of any age have the ability to opt out of automated decision-making processes~\cite{gdpr}. This research aids these ongoing regulatory discussions by providing previously missing quantitative evidence that interfaces employing autoplay \textit{can and do} affect users in measurable ways.

\section{Conclusion}
This research significantly enriches the ongoing discourse surrounding the tangible impact of platform design features on user consumption behaviors. 
By investigating the dynamics between platform design and user engagement patterns, this work serves as a compelling illustration of how ACDPs can wield influence over user behavior.
Specifically, we found that when participants disabled autoplay on Netflix, their session lengths and total watching times were reduced and their downtime between consecutive episodes increased.
Our findings contribute to a growing understanding of user interactions with entertainment platforms as potentially predatory agency exchanges~\cite{schaffner2023don} and add quantitative evidence to the ongoing discussion surrounding dark patterns and sovereignty over user decision-making.

\bibliographystyle{ACM-Reference-Format}
\bibliography{references}

\end{document}